\documentstyle[11pt,aaspp4]{article}

\slugcomment{to appear in  THE ASTROPHYSICAL JOURNAL}

\lefthead{JIANG ET AL.}
\righthead{THE INHOMOGENEOUS JET PARAMETERS IN AGNS}

\begin{document}

\title{THE INHOMOGENEOUS JET PARAMETERS IN ACTIVE GALACTIC NUCLEI}

\author{  D. R. JIANG, XINWU CAO, AND XIAOYU HONG    }
\affil{Shanghai Observatory, Chinese Academy of Sciences, Shanghai,
200030, China; djiang@center.shao.ac.cn}

\begin{abstract}
The K\"onigl inhomogeneous jet model is applied to investigate the
properties of the jets in Active Galactic Nuclei (AGNs). A sample of the
AGNs is collected, in which the measurements of the angular size and radio
flux density of the VLBI core, proper motion of the components in the jet,
and X-ray flux density are included. The inhomogeneous jet parameters are
derived with the same assumptions for all sources. A comparison among the
parameters of different types of sources in the sample is presented. It is
found that most of EGRET (Energetic Gamma-ray Experiment Telescope) sources
have higher Doppler factor $\delta $, larger Lorentz factor $\gamma $, and
smaller viewing angle $\theta $, when compared with the remaining sources in
the sample. The statistical analyses show that the derived Doppler
factor $\delta $ is strongly correlated with the observed 22 GHz brightness
temperature. Furthermore, there is a correlation between the relative $\gamma 
$-ray luminosity and the Doppler factor $\delta $. The implications of these
results are discussed.

\end{abstract}


\section{INTRODUCTION}
Superluminal motion has been observed in many active galactic nuclei
(AGN) with VLBI.
This provides strong evidence that the plasma in the jets
moves at relativistic velocity.
In the framework of the unified scheme, the
different classes of AGNs (such as radio galaxies, radio-loud quasars, and
blazars) can be interpreted as the same kind of sources but viewed at
different directions. Therefore the bulk velocity in the jets and the
viewing angles are two key parameters for our understanding the physics of
jets and discriminating models of AGNs.

The energetic $\gamma $-ray emissions from the AGNs offers an important clue
to understand the physics at work in the AGNs. Only one quasar (3C273) had
been detected in high energy $\gamma $-ray emission before the launch of the
Energetic Gamma-Ray Experiment Telescope (EGRET) on board the Compton
Gamma-Ray Observatory. So far, the EGRET has detected about 50 AGN (von
Montigny et al., 1995; Thompson et al. 1995). Some interesting results from
the observations are summarized as follows (von Montigny et al., 1995): (1)
The $\gamma $-ray energy flux in many of the sources is dominant over the
flux at lower energy bands. The typical isotropic apparent $\gamma $-ray
luminosities are in the range of $10^{45}-10^{49}$ ergs s$^{-1}$ . (2) Many of the sources exhibit rapid variability
on time scales from days to months, which implies that the size of emitting
region is of the order of the Schwarzschild radius of a black hole with $%
10^{10}$ M$_{\odot }$ under isotropic emission assumption. (3) Many active
galaxies relatively close to the Earth and some of the superluminal radio
sources have not been detected.

The proper motion measurements in the compact structures of the AGNs provide
useful information on the bulk motions of the emitting plasma. In the
framework of the relativistic beaming model and the synchrotron self-Compton
(SSC) model, the VLBI observations combined with the X-ray flux density
could be used to derive the Doppler boosting factor and some physical
quantities in the emitting regions of the AGNs. Marscher (1987) derived the
beaming parameters on the assumption of homogeneous spherical emission
plasma. Ghisellini et al. (1993) adopted Marscher's approach and obtained
the Doppler boosting factor $\delta $ for 105 sources. Readhead (1994)
suggested to estimate the value of equipartition Doppler boosting factor $%
\delta _{eq}$ using a single epoch radio data by assuming that the sources
are to in equipartition between the energy of radiating particles and the
magnetic field. G\"uijosa and Daly (1996) derived the $\delta _{eq}$ for the
same sample in Ghisellini et al. (1993). The advantage of homogeneous sphere
model is that the formalism is simple and the value of $\delta $ derived is
independent on the cosmology model. The component angular size and the flux
at the turnover frequency should be known in their calculation. In practice,
it is difficult to obtain this information, so one has to assume
that the VLBI observing frequency is the synchrotron self-absorption
frequency. In addition, the dependence of core size on the observing
frequency in some sources is inconsistent with the homogeneous spherical
assumption.

Blandford and K\"onigl (1979) and K\"onigl (1981) presented an inhomogeneous
relativistic jet model, in which both the flat spectrum characteristics of
some AGNs and the dependence of the core size on the observing frequency
could be well explained. K\"onigl's model involves more free parameters than
the homogeneous model, which limits its application. Some authors (Hutter
and Mufson 1986; Mufson et al, 1989; Unwin et al. 1994; Zensus et al. 1989
\& Webb et al. 1994) have explored the application of this kind of model to
obtain some physical parameters in some sources.

We assemble an AGN sample with both VLBI and X-ray observations available.
K\"onigl's inhomogeneous jet model is applied in our work to derive the
physical parameters of the jets. In $\S $ 2 we briefly describe the jet
model formalism and the approach. The sample and the results are described
in $\S $ 3 and $\S $ 4, respectively. The $\S $ 5 contains a short discussion.

\section{MODEL}

K\"onigl (1981) proposed an inhomogeneous jet model, in which the magnetic
field B(r) and the number density of the relativistic electrons ${n_e}(r,{%
\gamma _e})$ in the jet are assumed to vary with the distance from the apex
of the jet $r$ as $B(r)={B_1}(r/{r_1})^{-m}$ and ${n_e}(r,{\gamma _e})={n_1}%
(r/{r_1})^{-n}{\gamma _e}^{-(2\alpha +1)}$ respectively, and ${r_1}=1pc$. If
the bulk motion velocity of the jet is ${\beta }c$ (corresponding to a
Lorentz factor $\gamma $ ) with an opening half-angle $\phi $, and the axis
of the jet makes an angle $\theta $ with the direction of the observer, the
distance from the origin of the jet, $r({\tau _{\nu _s}}=1)$, at which the
optical depth to the synchrotron self-absorption at the observing frequency
$\nu _s$ equals unity, is given by equation (3) in K\"onigl (1981) as

\begin{equation}
{\frac{{r({\tau_{\nu_{s}}}= 1)}}{{r_1}}}=(2c_{2}(\alpha)r_{1}n_{1}\phi\csc
\theta)^{2/(2\alpha+5)k_m}(B_{1}\delta)^{(2\alpha+3)/(2\alpha+5)k_m}(\nu_{s}
(1+z))^{-1/k_m}.  
\end{equation}
Here $c_{2}(\alpha)$ is the constant in the synchrotron absorption
coefficient, $\delta$ is the Doppler factor, and $k_{m}=[2n+m(2\alpha+3)-2]
/(2\alpha+5)$. We use the projection of the optically thick region in the
jet as a measurement of the observed VLBI core angular size $\theta_d$,

\begin{equation}
\theta_{d}= {\frac{{r({\tau_{\nu_{s}}}= 1)\sin \theta}}{{\ D_{a}}}},
\end{equation}
where $D_a$ is the angular diameter distance of the source.

By integration of the emission from the optically thick region along the
jet, we obtain the radio flux of the core as

\begin{displaymath}
s(\nu_{s})={\frac{1}{{4\pi {D_{a}^{2}}}}} {\frac{{c_{1}(\alpha)}}{{%
c_{2}(\alpha)}}}B_{1}^{-1/2}\left({\frac{\delta}{{1+z}}}\right)^{3} 
\left [ {\frac{{\nu_{s}(1+z)}}{{\delta}}}\right
]^{5/2}\int\limits _{0}^{r(\tau_{\nu_s}=1)}2\left({\frac{r}{{r_{1}}}}%
\right)^{m/2}\varphi r \sin \theta dr%
\end{displaymath}
\begin{equation}
={\frac{{r_{1}^{2}\phi\sin\theta}}{{(4+m)\pi D_{a}^{2}}}} {\frac{{\
c_{1}(\alpha)}}{{c_{2}(\alpha)}}} B_{1}^{-1/2}\nu_{s}^{5/2} \left({\frac{
\delta}{{1+z}}}\right)^{1/2}\left({\frac{{r({\tau_{\nu_{s}}}= 1)}}{{r_1}}}
\right)^{(4+m)/2}, 
\end{equation}
where $\nu _s$ is the VLBI observing frequency, and ${c_1}(\alpha )$ and ${%
c_2}(\alpha )$ are the constants in the synchrotron emission and absorption
coefficients, respectively.

Equation (13) in K\"onigl's work gives the X-ray flux density estimation
from an unresolved jet. We adopt his expression in the frequency region ${%
\nu _c}>{\nu _{cb}}(r_M)$, where $r_M$ is the smallest radius from which
optically thin synchrotron emission with spectral index $\alpha $ is
observed (K\"onigl 1981).

The proper motion observed with VLBI could be converted to the apparent
transverse velocity $\beta _{app}$ by using the Friedmann cosmology. The
apparent transverse velocity $\beta _{app}$ is related to the bulk velocity
of the jet ${\beta }c$ and viewing angle $\theta $,

\begin{equation}
\beta_{app}={\frac{{\beta\sin\theta}}{{1-\beta\cos\theta}}},  
\end{equation}
if the viewing angle $\theta $ is available.

The above equations, in conjunction with equation (13)
in K\"onigl(1981), can well describe the relation between the physical
parameters of the inhomogeneous relativistic jet model and the observational
results. The parameters of an inhomogeneous jet could be derived from both
VLBI and X-ray observations given the three parameters $\alpha $, $m$, $n$,
and the relation between the opening half angle $\phi $ and the Lorentz
factor $\gamma $. Three parameters $\alpha $, $m$ and $n$ are related to $%
\alpha _{S1}$, $\alpha _{C2}$ and $k_m$ as following:

\begin{equation}
\alpha _{S1}={\frac 52}-{\frac{{4+m}}{{2k_m}}}, 
\end{equation}
\begin{equation}
\alpha_{C2}=\alpha+{\frac{{(1+\alpha)m+2n-4}}{{7m-4}}}  
\end{equation}
and

\begin{equation}
k_{m}={\frac{{2n+m(2\alpha+3)-2}}{{2\alpha+5}}},  
\end{equation}
where $\alpha _{S1}$ is the spectral index of the VLBI core at radio band, $%
\alpha _{C2}$ is the spectral index at X-ray band ($S_\nu \propto \nu
^{-\alpha }$) and $k_m $ relates the dependence of the core angular size
on the observing frequency ($\theta _d\propto \nu _{ob}^{-{1/{k_m}}}$).

In principle, three parameters $\alpha$, $m$ and $n$ could be constrained by
the observable quantities $\alpha_{S1}$, $\alpha_{C2}$ and $k_m$. Many
workers have tried to explore the mean value of the spectral index.
Padovani et al. (1997) suggested that the mean value of the spectral index
in the X-ray band of the flat-spectrum radio quasars is
$\langle\alpha_{x}\rangle \sim1$. Brunner et al. (1994) obtained
$\langle\alpha_{x}\rangle\sim0.6$ for radio
loud quasars. Lamer et al. (1996) found $\langle\alpha_{x}\rangle\sim 1.30$
for BL Lac objects. Padovani and Urry (1992) used $\langle\alpha_{S1}\rangle
\sim-0.1$ for flat spectrum radio quasars.
For BL Lac objects, $\langle\alpha_{S1}\rangle
\sim-0.3$ is given by Padovani (1992). In the cases where there are
multi-frequency
VLBI observations, $k_{m}\sim1$ is found.

Unwin et al.(1994) found $\alpha =0.6$, $m=1.5$ and $n=1.4$ for the conical
jet model in 3C345. The values of $k_m$, $\alpha _{S1}$ and $\alpha _{C2}$
in that case are $1.145$, $-0.1$ and $0.78$, respectively. These values
agree with the statistical values. Hutter and Mufson (1986) expected $m=1$
and $n=2$ for a free jet. Webb et al. (1994) derived $m=0.85-1.15$, $%
n=1.77-2.4$ for 3C345.

The projection of the opening half-angle $\phi _{ob}=\phi /\sin \theta $ is
a measurable quantity. But this kind of the information is available only in
a few sources. To derive the parameters of the inhomogeneous jet, we have to
make some simplified assumptions. Blandford and K\"onigl (1979) suggested $%
\phi \leq 1/\gamma $. Some authors (Hutler and Mufson 1986, and Mufson et
al. 1988) adopted the assumption $\phi =1/\gamma $ in their application of
this model. Marscher (1987) argues that $\tan \phi =(\sqrt{3}\gamma )^{-1}$ for
a free jet.

In our calculation, we assume $\alpha=0.75$, $m=1$, $n=2$ and the opening
half-angle $\phi=1/\gamma $ in Model A and $\alpha=0.6$, $m=1.5$, $n=1.4$
and $\phi=1/\gamma $ in Model B. The four independent variables $n_1$, $B_1$,
$\beta$, and $\theta$ can be derived from Equations (2)$-$(4) and
K\"onigl's Equation (13) (K\"onigl 1981). Thus, we obtain simultaneously
the values of Lorentz factor $\gamma$, viewing angle $\theta$, and
Doppler factor $\delta$.

Equation (1) of Ghisellini et al. (1993),

\begin{equation}
\delta=f(\alpha)F_{m}\left[{\frac{{\ln({\nu_b}/{\nu_m})}}{{\ {F_{x}} {{\theta%
}_{d}^{6+4\alpha}} \nu_{x}{{\nu}_{m}^{5+3\alpha}} } }}\right ]
^{1/(4+2\alpha)}(1+z), 
\end{equation}
is used to compare the results of the homogeneous sphere model, where $F_x$
is the X-ray at frequency $\nu _x$, $F_m$, $\theta _d$ are the radio flux
density and the angular size of the core at the turnover frequency $\nu _m$.
The VLBI observing frequency is assumed to be $\nu _m$. We assign the
homogeneous sphere model with $\alpha =0.75$ as Model C . The values ${H_0}%
=75~kms^{-1}Mpc^{-1}$ and ${q_0}=0.5$ are used throughout this work.

\section{ SAMPLE}

We have searched the literature for objects which have relevant
data, such as the radio flux density and the size of the core, the proper
motion and X-ray flux density . The VLBI core and X-ray data have been
presented by Ghisellini et al. (1993), while Vermeulon and Cohen (1994) have
compiled the proper motion data. The selection criterion of our sample is
that all sources have VLBI measurements of proper motion of outflowing
plasma. A total of 52 sources were chosen after a careful literature search,
in which 17 sources are detected EGRET $\gamma$-ray sources. The
observational data for the sample are presented in Table 1, which gives the
redshift (z), VLBI core size ($\theta _d$), and radio flux ($S_c$) at the
frequency $\nu _s$, 1 keV X-ray flux density ($S_x$) and the proper motion ($%
\mu _{app}$) with necessary references. The redshift of 0716$+$714 is not
available, and a value of 0.3 is assumed in the calculation. For 1823+568,
only the 2 keV X-ray flux density is given, and the 1 keV X-ray flux density
was derived by assuming a spectral index of 1.30, which is the average value for BL Lacs.
We assume that all the observed X-ray flux density is attributed to the SSC
emission in the derivation, which will result in some uncertainties. Also
the X-ray observations are not contemporaneous with the VLBI observations,
and this will introduce some uncertainties too. However our calculations show
that the value of X-ray flux density is not sensitive to the derived
parameters. Another uncertainty of data is the proper motion. We assume the
fastest one when there are more than one moving components. Even though it is not
quite certain that the observed proper motion could represent the real
information of the core, we think that the observed proper motion is a good
approximation.

\section{RESULTS}

Using the method described in $\S $ 2 the jets' parameters of all the
sources in our sample are derived. The results are shown in Tables 2A and 2B
for above mentioned models A and B, respectively. We note that there is no
big difference between the results from these two models, so we will discuss
only model A when comparing to the homogeneous model (model C).

\subsection{\it The distribution of Doppler factor and Lorentz factor}

In Figure 1 we show the distribution of the derived Doppler factor $\delta $
for the 17 EGRET sources and the 45 remaining sources in our sample. The
EGRET sources have higher values of $\delta $, except three BL Lacs: 0716$+$%
714, 1101$+$384(Mrk421), and 1219$+$285. The distribution of the Lorentz
factor $\gamma $ are plotted in Figure 2. Similarly, it is found that the
EGRET sources have relatively high values of $\gamma $. From Figures 1 and 2
we note that there are some objects with high values of $\delta $ or $\gamma 
$ have not been measured $\gamma $-ray radiation by EGRET. Compared with the
other EGRET sources, three BL Lacs 0716$+$714, Mrk421, and 1219$+$285 have
relatively low values of both $\delta $ and $\gamma $. The BL Lac object
Mrk421 is a special one because of its detection as an TeV source as well as
EGRET source. The redshift of object 0716$+$714 is not available, with an
assumed 0.3, which would probably increase the uncertainty of its derived
parameters. These three objects show interesting characters.

Figure 3a is the plot of Lorentz factors $\gamma $ for all the objects in
our sample with $\delta >\delta _c$, where $\delta _c=4.5$ for model A. The
14 EGRET sources show large values of Lorentz factor $\gamma $ with $\gamma
\geq $ 10. Similar phenomena are shown for model B, where $\delta _c=3.5$.
Almost all of the remaining objects with $\delta >\delta _c$ show relatively
lower Lorentz factors $\gamma $ except the source 1308$+$326. Figure 3b is
the results of model C which has $\delta _c=2.5.$

The statistical information of the Doppler factor $\delta $, Lorentz factor $%
\gamma $, and the viewing angle $\theta $ for the different classes of the
sources in our sample are listed in Table 3. In Figure 4 the Lorentz factor $%
\gamma $ vs. the viewing angle $\theta $ for all sources is depicted.

\subsection {\it The distribution of turnover frequency}

In our model the observed turnover frequency $\nu _{sM}$ of the radio
continuum could be naturally derived. If there are enough dedicated 
measurements on the turnover frequency, it will offer an effective
examination on our model. Estimates of the synchrotron self-absorption
frequency for some sources (Bloom et al., 1994; Reich et al., 1993)
seem to be compatible with our results. We depict the distribution of the turnover frequency in
the sources' rest frame $(1+z)\nu _{sM}$ in Figure 5. The EGRET sources
have relatively higher turnover frequencies measured in the sources' rest frame
compared with the other sources in our sample. This could be the natural
conclusion drawn from the fact that the EGRET sources have high values of
Doppler factor ($\S $ 4.1). No obvious difference is found in the intrinsic
turnover frequencies ($\nu _{sM}(1+z)/\delta $) between the EGRET and
the rest of the sources in our sample.

\subsection {\it  Correlation of brightness temperature $(1+z)T_{b}$ in the source's
rest frame with Doppler factor $\delta$}

In Figure 6a and 6b we plot the brightness temperatures in the sources' rest
frame $(1+z)T_b$ vs. the Doppler factor $\delta $ derived from model A and
C, respectively. The observed 22GHz brightness temperatures $T_b$ are taken
from Moellenbrock et al. (1996). The total number of sources is 31, within
which 13 sources only have low limits on the brightness temperature. There
are significant correlations (at 99.9\% for both model A and B) between the
brightness temperature in the source's rest frame and the Doppler factor
derived from model A and B, respectively. In calculating the correlation,
the source 0016$+$731 is not included, the brightness temperature of this
source is only given a lower limit. The best correlation for the data derived from model A and B are: $\log
(1+z)T_b=10.98+0.84\log \delta $ and $\log (1+z)T_b=11.00+0.93\log \delta $,
respectively. A less significant correlation is found for model C:
$\log (1+z)T_b=11.22+0.45\log \delta $   (a
correlation coefficient $r=0.39$, at 98\% level). The correlation derived
from our model seems to agree with the prediction: $(1+z)T_b\propto \delta $.

The correlation between the values of the Doppler factor $\delta $ derived
from models A and C is also examined as shown in Figure 7. A good
correlation is found, which means the statistical behavior of the Doppler
factor derived from homogeneous sphere model has no significant difference
from our model. Nonetheless, the values of the Lorentz factor $\gamma $
derived in the homogeneous sphere model for some sources would be as high as
several hundred, which is mainly due to extremely low estimate on the
Doppler factor for these sources (Ghisellini et al. 1993). The proper motion
data are not taken into account in the derivation of the Doppler factor $%
\delta $ in homogeneous sphere model. In our model, the proper motion
information is used in the derivation of both the Doppler factor $\delta $
and Lorentz factor $\gamma $. We suggest that the homogeneous sphere model
could be used especially for the sources without the data of proper motion.

\subsection {\it Correlation of the relative $\gamma$-ray luminosity with the
Doppler factor $\delta$}

Recently, about fifty AGN have been detected high energy $\gamma $-ray
emission by EGRET(von Montigny et al., 1995; Thompson et al. 1995). 17 EGRET
sources are listed in our sample. We present the correlation of the relative 
$\gamma $-ray luminosity $L_\gamma $ with the Doppler factor $\delta $ in
Figure 8a. The relative $\gamma $-ray luminosity $L_\gamma $ is simply
defined by $L_\gamma ={d_L^2}F$, where $d_L$ is the luminosity distance, $F$
is the maximum photon flux in $\gamma $-ray range taken from von Montigny et
al. (1995) and Thompson et al. (1995). The best correlation for the data
derived from models A and B are: $\log L_\gamma =-1.80+2.45\log \delta $ (at
99.9\% level) and $\log L_\gamma =-1.59+2.52\log \delta $ (at 99.9\% level),
respectively. We find the correlation, $\log L_\gamma =-0.94+1.58\log \delta 
$ (at 99.9\% level), for model C, which is plotted in Figure 8b.

\section{DISCUSSION}

We summarize the main results obtained in the previous sections as follows:

1. The values of the Doppler factor $\delta $ of the EGRET sources are
higher than those of the non EGRET sources in our sample except three BL
Lacs: Mrk421, 0716$+$714, and 1219$+$285, though some sources other than the
EGRET sources also have high $\delta $. The values of Lorentz factor $\gamma 
$ of the EGRET sources show similar behavior.

2. The EGRET sources except the three BL Lacs, have both large values of
the Lorentz factor $\gamma$ ($\geq 10$) and $\delta$ ($\delta\geq\delta_{c}$%
, $\delta_{c}=4.5, 3.5$, for models A and B, respectively), which seems to
be the significant difference between the EGRET and the remaining sources in the
sample.

3. The derived turnover frequency $(1+z)\nu_{sM}$ measured in the sources'
rest frame are higher for the EGRET sources than the rest sources in the
sample, though there still some sources other than the EGRET sources have
high turnover frequencies.

4. The BL Lacs have a similar mean Lorentz factor $\gamma $ with the
core-dominated quasars, but the mean viewing angle $\theta $ of the BL Lacs
is slightly larger than that of the core-dominated quasars. The
lobe-dominated quasars have a large mean viewing angle $\sim 40^{\circ }$,
while for the core-dominated quasars it is $\sim 10^{\circ }$.

5. There are significant correlations between the brightness temperature in
the sources' rest frame $(1+z)T_b$ and the Doppler factor $\delta $ for both
Models A and B. No similar significant correlation is found for Model C.

6. The significant correlations of the relative $\gamma$-ray luminosity $%
L_{\gamma}$ are found with the Doppler factor $\delta$ for both model A and
B.

There are 17 EGRET sources included in our sample. Except three BL Lacs,
all of the remaining 14 EGRET sources show high value of $\delta $ and $\gamma $%
. The mean viewing angle of 14 EGRET sources are 4.9$^\circ$ and 5.7$^\circ$ for
Model A and B, respectively. We have not seen any difference on the intrinsic physical
properties between these two types of sources from our statistical results.
These suggest that the $\gamma $-ray emissions from the AGN are mainly
due to the beaming effects. Only the sources with high Doppler factor
$\delta $, which are strongly beamed to us, are detected at $\gamma $-ray
energies. For the synchrotron self-Compton model of $\gamma $-ray
production, the Doppler factor $\delta $ rather than Lorentz factor $\gamma $
of the bulk motion of the jet plays crucial roles on the observed $\gamma $%
-ray flux.
Sikora, Begelman \& Rees (1993, 1994) have proposed that the
$\gamma$-ray emission originates in a jet as a product of inverse Compton
scattering of relativistic electrons and seed photons produced externally
to the jet. Moving in a homogeneous photon 'bath' (produced by material
scattering photons from the disk, or by the broad-line region), the
jet would 'see' this radiation energy density amplified by a factor
$\gamma^2$. Therefore in this model the Lorentz factor $\gamma$ together
with Doppler factor $\delta$ determine the $\gamma$-ray radiation. 
The fact that some sources with high Doppler factor $\delta $ and
low Lorentz factor $\gamma $ have not been detected the $\gamma $-ray emission
might imply that the seed photons are from somewhere outside the jet, if the 
$\gamma $-ray detected by EGRET is due to the inverse Compton scattering of
lower energy photons up to $\gamma $-ray energies by beamed relativistic
electrons.
Thus, the sources with low Lorentz factor cannot produce
sufficient $\gamma $-rays through inverse Compton scattering of the external
soft photons.

The source 1308$+$326 has both high $\delta $ and $\gamma $ similar to the
EGRET sources in the sample but it is non EGRET source. This might be due to
the measurements on the value of proper motion. The overestimation on the
proper motion of the source would lead to the overestimation of its $\delta $
and $\gamma $ . The further VLBI measurement on the proper motion of this
object is necessary. The other possibility is that the object is a $\gamma $%
-ray source just in the quiescent state and therefore has not been detected
by EGRET.

We cannot see significant differences on the mean values of $\delta $, $%
\gamma $, and $\theta $ of the BL Lacs from that of the all sources in the
sample (Table 3). The mean value of the Lorentz factor $\gamma $ of the BL
Lacs is similar to that of the core-dominated quasars and the mean viewing
angle of BL Lacs is slightly larger than that of the core-dominated quasars.
The mean values of Doppler factor show that the core-dominated quasars are
more beamed than the BL Lacs. The results obtained here seem not to agree
with the previous suggestion that the viewing angle of BL Lacs are smaller.
Three BL Lacs Mrk421, 0716$+$714 and 1210$+$285 are quite special, which
have relatively low Lorentz and Doppler factors, and large viewing angle,
but are detected at $\gamma $-ray energies. One possible reason might be
that the BL Lac objects have a different radiation mechanism from the other
AGNs. Urry (1994) suggested that the X-ray emission from the BL Lacs are due
to the synchrotron radiation instead of the synchrotron self-Compton
radiation. Thus, there may be some problems in using a single model to
describe all AGNs.

It is shown in Table 3 that the core-dominated and lobe-dominated quasars
have rather different mean values of viewing angle. These two types quasars
may be the same phenomenon, but seen at different viewing angles, which is
consistent with the previous results (Ghisellini et al. 1993) and the
unified scheme.

The correlation between the brightness temperature in the source's rest
frame $(1+z)T_b$ and the Doppler factor $\delta $ derived in our model
suggest that the derived values of beaming parameters are a good
approximation. As a comparison the correlation is less significant
for the homogeneous sphere model. However, the homogeneous sphere
model is useful to estimate the Doppler factor, especially for the objects
without proper motion data, since the derived Doppler factor from the
homogeneous sphere model is in general compatible with that from our
inhomogeneous jet model (see Figure 7). The correlation of the relative $%
\gamma $-ray luminosity $L_\gamma $ with the Doppler factor $\delta $
presented in Figure 8 strongly suggests that the $\gamma $-ray emission from
the AGN are beamed, though the detailed mechanism for $\gamma $-ray emission
is still not clear.

Two sets of parameters $\alpha $, $m$, and $n$ are adopted in the model
calculations and we adopted the same values of these parameters for all
sources in one model calculation. In practice, the sources may have
different values of parameters $\alpha $, $m$, and $n$, and, in principle,
these parameters could be constrained by the observable quantities $\alpha
_{S1}$, $\alpha _{C2}$, and $k_m$. Unfortunately, the information is only
found for a few cases through multi-frequencies VLBI observations. Further
high resolution multi-frequencies VLBI observations would be helpful to
improve our model calculations.

\acknowledgments
We thank the referee for his helpful comments 
on the manuscript. The support from Pandeng
Plan is gratefully acknowledged. XC thanks the support from Shanghai
Observatory, China Post-Doctoral Foundation, and NSFC.

\clearpage

\figcaption{
The distributions of the Doppler factor $\delta$ for the EGRET
sources (solid) and the rest of the sources in the sample (dotted) derived from
Models A (a) and C (b), respectively.}

\figcaption{
The distributions of the Lorentz factor $\gamma$ for the EGRET
sources (solid) and the rest of the sources in the sample (dotted) derived from
Models A (a) and C (b), respectively. }

\figcaption{
The Lorentz factor $\gamma$ for the sources with Doppler factor $%
\delta> \delta_{c}$ in the sample corresponding to Models A (a) and C (b),
respectively. {\it Triangles}: EGRET sources; {\it circles}: the rest
sources.}

\figcaption{
The Lorentz factor $\gamma$ vs. the viewing angle $\theta$ for all
sources in the sample derived from Model A. {\it Triangles}: EGRET sources; 
{\it circles}: the rest sources. The dashed line represents $%
\gamma=1/\sin\theta$.}

\figcaption{
The distributions of the turnover frequency $\nu_{sM}(1+z)$ in the
source's rest frame for the EGRET sources (solid) and the rest sources in
the sample (dotted) derived from Models A (a) and C (b), respectively. }

\figcaption{
The intrinsic brightness temperature in the source's rest frame vs.
the Doppler factor $\delta$ derived from Models A (a) and C (b),
respectively. Arrow symbols indicate the low limits on the brightness
temperature.}

\figcaption{
The Doppler factor $\delta$ derived from Model A vs. that from Model
C. {\it Triangles}: EGRET sources; {\it circles}: the remaining of
sources.   }

\figcaption{
The relative $\gamma$-ray luminosity vs. the Doppler factor $\delta$
derived from Models A (a) and C (b), respectively. }

\end{document}